\documentclass[a4paper,journal,twoside]{IEEEtran}
\usepackage{amsmath,amsfonts}
\usepackage{algorithmic}
\usepackage{algorithm}
\usepackage{array}
\usepackage[caption=false,font=normalsize,labelfont=sf,textfont=sf]{subfig}
\usepackage{textcomp}
\usepackage{stfloats}
\usepackage{url}
\usepackage{verbatim}
\usepackage{graphicx}
\usepackage{cite}
\usepackage{color}
\usepackage{siunitx}
\usepackage{alltt}
\usepackage[table]{xcolor}

\usepackage[acronym]{glossaries}

\newacronym{ecu}{ECU}{Electronic Control Unit}
\newacronym{can}{CAN}{Controller Area Network}
\newacronym{sc}{SC}{Security Channel}
\newacronym{sz}{SZ}{Secure Zone}

\newcommand \Tsc {\mathrm{T}_\mathrm{SC}}
\newcommand \Rsc {\mathcal{R}_\mathrm{SC}}

\usepackage{lipsum}%
\usepackage[pscoord]{eso-pic}
\newcommand{\placetextbox}[3]{%
  \setbox0=\hbox{#3}
  \AddToShipoutPictureFG*{
    \put(\LenToUnit{#1\paperwidth},\LenToUnit{#2\paperheight}){\vtop{{\null}\makebox[0pt][c]{#3}}}%
  }%
}%

\begin{document}
\placetextbox{0.5}{1}{This work has been submitted to the IEEE for possible publication. Copyright may be}
\placetextbox{0.5}{0.985}{transferred without notice, after which this version may no longer be accessible.}

\title{Robust Multicast Origin Authentication in MACsec and CANsec for Automotive Scenarios}

\author{
    \IEEEauthorblockN{Gianluca Cena, \textit{Senior Member, IEEE}, Lucia Seno, and Stefano Scanzio, \textit{Senior Member, IEEE}}
}

\markboth{}%
{Cena \MakeLowercase{\textit{et al.}}: Robust Multicast Origin Authentication in MACsec and CANsec for Automotive Scenarios}

\maketitle

\begin{abstract}
Having everything interconnected through the Internet, including vehicle onboard systems, is making security a primary concern in the automotive domain as well.
Although Ethernet and CAN XL provide link-level security based on symmetric cryptography, they do not support origin authentication for multicast transmissions. 
Asymmetric cryptography is unsuitable for networked embedded control systems with real-time constraints and limited computational resources.
In these cases, solutions derived from the TESLA broadcast authentication protocol may constitute a more suitable option.

In this paper, some such strategies are presented and analyzed that allow for multicast origin authentication, also improving robustness to frame losses by means of interleaved keychains.
A flexible authentication mechanism that relies on a unified receiver is then proposed, which enables transmitters to select strategies at runtime, to achieve the best compromise among security, reliability, and resource consumption.
\end{abstract}


\section{Introduction}
\label{sec:introduction}
\IEEEPARstart{S}{ecurity} issues are becoming increasingly relevant for networked embedded control systems, particularly in the automotive domain because of the increase in vehicle intra and inter-connectivity. Since the 1990s, digital networks, and in particular the \gls{can} \cite{ISO11898}, have been universally adopted for the communication among \glspl{ecu}. 
Although providing many advantages, digital networks unavoidably increase the vehicle attack surface. 
As an example, by connecting  devices programmed to emulate key fobs for emergency start (available on the dark web \cite{2023-Tindell}) to the CAN bus, thieves can unlock a car doors and turn the engine on \cite{2023-InjAttack}.
This kind of attacks, known as CAN injection, requires physical access to the car in order to plug the malicious hardware to the in-vehicle network.

Luckily, CAN injection is natively prevented, or at least consistently hindered, by next generation in-vehicle networks.
As an example, automotive Ethernet, including IEEE 802.3cg (10Base-T1S) \cite{IEEE8023CG}, provides MAC-level security (integrity and confidentiality) by means of the IEEE 802.1AE (MACsec) standard \cite{IEEE8021AE}.
Likewise, the upcoming (third) version of CAN, known as CAN XL \cite{DS610-1}, is planned to include its own security layer, termed CANsec \cite{DS613-2}, which stems from MACsec but explicitly targets the automotive scenario.
Both MACsec and CANsec rely on \textit{symmetric} cryptography, and are simple enough so that they can be realistically included in next-generation \glspl{ecu} at modest cost, complementing end-to-end security. 
When the above solutions are in place, physically tampering the (local) in-vehicle network does not grant intruders control on the car, unless the keys used for link-level cryptography are known.

Cars are however evolving and progressively becoming \textit{connected} cars \cite{2015-ITS, 2020-Acc-Chowd,2022-ITS}.
Besides vehicle-to-everything (V2X) communications, which enable features like platooning and coordination with traffic lights, most of today's cars connect to the Internet to download contents, i.e., the user can interact with the dashboard to obtain data from third parties, including traffic information, maps for navigation, firmware updates, and specific apps providing enhanced driving functions \cite{2014-JIoT}.
Some carmakers are even proposing that software modules implementing specific add-on functions (e.g., park assist) can be bought online by the client after the car purchase.
Moreover, vehicle on-board diagnostics is nowadays increasingly often carried out by means of dongles connected to OBD-II port, which communicate via Wi-Fi/Bluetooth with smartphones running suitable apps for troubleshooting \cite{2017-IEEE-Sens-J}.
Data loggers that communicate directly over LTE/5G also exist, e.g., to remotely track and manage fleets of vehicles \cite{2022-TITS-OBD}.

Connectivity to the outside world further increases vehicle attack surface. As an example, software vulnerabilities can be exploited by attackers to inject malicious code in ECUs connected to the Internet.
A compromised ECU is then able to forge and send messages to other ECUs, e.g., to suddenly activate the brakes or increase the speed beyond safe levels.
The impact of this kind of attacks (vehicle borne threats \cite{HVM}) is potentially far more dangerous than car injection, as a multitude of vehicles could be silently infected over time (months) and the injected malware might force all of them to simultaneously perform unwanted actions, causing severe damages and even casualties. 
It is easy to imagine what a disastrous impact may have thousands of vehicles suddenly getting out of control. 

Solutions like MACsec and CANsec can mitigate such attacks.
In fact, the control system of a vehicle can be partitioned into distinct \glspl{sz} so that, e.g., the infotainment system is not enabled to interact with the brakes, simply because there is no \gls{sc} that foresees messages going in that direction.
Unfortunately, this is true only if communications are point-to-point, that is, if \glspl{sc} only involve two devices.
In practice, almost all current designs exploit multicast communication in order to save bandwidth.
For instance, the angular speed of the wheels is concurrently consumed by the Anti Brake-locking System (ABS), the dashboard for the speedometer, and possibly the multimedia system for automatically adjusting the volume based on the car speed.
Therefore, a multimedia system that has been compromised because, e.g., a fake codec was downloaded by a careless user from a fraudulent site, may impersonate the wheels and trick the ABS, undermining its operations.
Depending on the car model, other critical attacks may be delivered as well.

This problem can be faced in several ways: first, by replacing any secure multicast  connection with a bunch of unicast ones.
Besides increasing bandwidth consumption, doing so requires every producer to be aware of all its consumers.
Alternatively, the network can be partitioned by switches that incorporate firewall functionality.
This solution permits to block many attacks, depending on the ports where ECUs are connected, but is more expensive than buses and introduces additional delays, which explains why not every carmaker is willing to embrace it.
Another possibility is to adopt \textit{asymmetric} cryptography (i.e., certificates), which permits receivers to determine whether or not a certain message is authentic, i.e., to identify with certainty if the originator is what they expect.
However, the related computational complexity is much higher than symmetric cryptography \cite{Crypto}, and thus securing this way every real-time message cyclically exchanged on the network is unfeasible for the microcontrollers typically employed in embedded systems.

In this paper, a solution is explored for providing robust multicast origin authentication in the automotive and  embedded scenarios that relies on Lamport's keychains \cite{1979-Lamport} and resembles the Guy Fawkes protocol \cite{1998-SIGOPS}, from which the Timed Efficient Stream Loss-tolerant Authentication (TESLA) \cite{2000-SecPri} derives.
Similar approaches, targeting application domains with very different characteristics, were described in \cite{2016-NAVI}, which introduces the Galileo Open Service Navigation Message Authentication (OSNMA) service for the global navigation satellite system.
Interesting attempts were made to port TESLA to CAN \cite{2013-TII-Groza}, but they mostly focus on overcoming CAN limitations concerning payload size ($\SI{8}{B}$ at most), which have been completely solved by CAN XL.
A very interesting solution, named Inf-TESLA, is described in \cite{2016-ICTsys}, subsequently enhanced in \cite{2022-ACC-InfTesla}, which improves robustness to losses by using two alternating displaced keychains.
Our proposal, we term Tesla-based Reliable Unified-scheme with Dual-Interleaved keychain (TRUDI), defines a class of origin authentication strategies that resemble Inf-TESLA.
However, consistent differences exist concerning the following aspects:
\begin{enumerate}
    \item it relies on a simplified TESLA version with immediate authentication and delivery, which best suits real-time operation of most automotive and embedded systems;
    \item it is designed to offer utmost reliability against frame losses by increasing synchronization opportunities, yet keeping communication overhead as low as possible;
    \item it is an add-on for existing solutions like MACsec and CANsec, and only provides origin authentication;
    \item it defines a unified receiver architecture that gives the transmitter flexibility in deciding authentication strategy at runtime, to offer the best compromise among security, reliability, and resource consumption.
\end{enumerate}

The paper is structured as follow: Section~\ref{sec:Secure} briefly introduces link-level security and multicast origin authentication, while Section~\ref{sec:Origin} defines a number of solutions that rely on operating principles similar to TESLA.
Section~\ref{sec:Unified} presents our unified authentication strategy and shows how it permits implementing the whole range of strategies. Finally, some conclusions are drawn in Section~\ref{sec:Conc}.

\section{Multicast authentication in automotive}
\label{sec:Secure}
\begin{figure}[tb]
    \centering
    \includegraphics[width=0.9 \columnwidth]{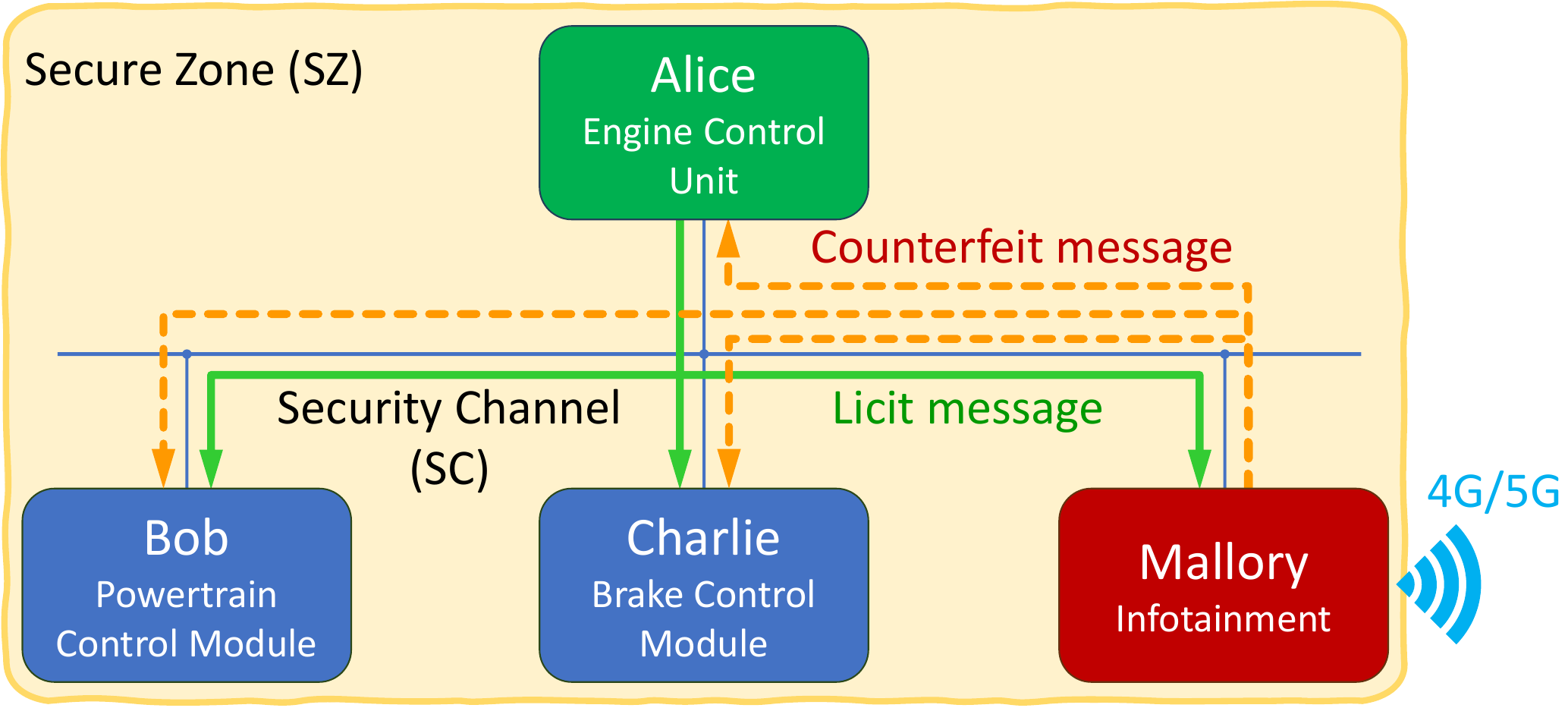}
    \caption{Masquerade attack by a compromised receiver in the SC.}
    \label{fig:alice}
\end{figure}

Both MACsec and CANsec rely on the Advanced Encryption Standard (AES) \cite{2001-AES}, and ensure message integrity and confidentiality.
Moreover, by resorting to message numbering (freshness values), they also offer protection against replay attacks \cite{1993-CSFW}, which could exploited, e.g., in injection attacks to bypass the need for the encryption key.
However, when multicast transmissions are considered, above solutions fail to guarantee origin authentication.
In fact, even if symmetric cryptography is exploited, all members of a \gls{sc} know the \textit{shared encryption key} $K_\mathrm{SC}$ and, if compromised, they can pretend to be the only node that is entitled to transmit in the \gls{sc} (termed $\Tsc$).
This is known as \textit{masquerade attack}, and is illustrated in Fig.~\ref{fig:alice}: Mallory (e.g., a car infotainment system that has been tampered by an attacker during a firmware update over the air) is enabled to send malicious messages that will be considered licit, i.e., sent by Alice ($\Tsc$), by Bob and Charlie.
Alice might be able to understand that something is wrong, but in general she will be unable to counteract the attack timely and effectively, since in the \gls{sc} she is indistinguishable from Mallory.
At best, she can react to an ongoing attack by broadcasting an alert to notify Bob and Charlie \cite{2017-ISCAS}, but doing so creates another vulnerability: in fact, such alerts can be exploited to deliver denial-of-service (DoS) attacks.

What is sought for networked embedded control systems like modern vehicles, where some nodes are connected to the Internet and may be compromised, is a robust yet efficient mechanism for message origin authentication that prevents masquerade attacks in the  \gls{sc}, achieving at the same time a balanced trade-off among security, reliability (two of the most relevant dependability attributes in these contexts, among those defined by Laprie et Al. \cite{TDSC04-Laprie}), and resource consumption (i.e., CPU time, memory usage, and network bandwidth).
Since security solutions for above systems already exist, e.g., MACsec and CANsec, enabling inexpensive and backward-compatible implementations requires that origin authentication be defined as an add-on, which can be layered on top of them (for instance, as a library to be linked to applications).

The most relevant performance indicator for any origin authentication mechanism is its ability to detect masquerade attacks.
We define the \textit{security level} as one minus the probability that a counterfeit message is taken as valid.
This event corresponds to a false positive for the message validation function $V(\cdot)$, whose aim is to verify the message source.
Conversely, the \textit{reliability level} is related to the probability that a licit message is discarded by the origin authentication mechanism (which leads to a false negative).
Having the network occasionally losing some messages is inevitable, but missing them because of a poorly designed authentication approach shall be definitely avoided.
Both false positives and false negatives may lead to dangerous situations, where vehicles, passengers, and third parties are exposed to severe risks: in the former case the attacker gains control of some subsystem, whereas in the latter communication on the \gls{sc} can be impaired for a while, preventing control loops from working correctly.
It is worth pointing out that, since we are dealing with systems with limited transmission bandwidth (typically in the order of $\SI{10}{Mb/s}$ \cite{2023-Decker}) and computational power (e.g., $\SI{32}{b}$ ARM CPUs), overheads (additional protocol control information needed for origin authentication) must be kept as low as possible, while preserving the target security and reliability levels.

\subsection{Notation and keychain-based authentication}
In the following we mostly reprise the notation in \cite{2000-SecPri}.
The stream of application-layer messages sent on a given \gls{sc} is denoted as $(M_j)_{j \geq 1}$.
Every message $M_j$ is encapsulated into a data-link frame $P_j$.
As we are dealing with link-layer security, we prefer not using the term ``packet'', which usually indicates network-layer datagrams. 
Moreover, since networked embedded control systems typically do not have any particular requirements about confidentiality, message encryption will not be considered, i.e., $M_j$ is assumed to be included in $P_j$ in cleartext. 
However, the proposed reasoning can be easily extended to include such data protection mechanisms as well.

A frame $P_j$ protected through MACsec or CANsec, e.g., by exploiting the Galois Message Authentication Code (GMAC) of the AES \cite{2007-dworkin}, is modeled as $P_j = \langle D_j, \mathrm{MAC}(K_\mathrm{SC}, D_j) \rangle$, where $\mathrm{MAC}(\cdot)$ denotes the \textit{message authentication code}, also known as integrity check value (ICV), computed by the algorithm based on the shared key $K_{SC}$, while $D_j = \langle L_j, I_j, M_j \rangle$ collectively denotes all the frame information protected by the MAC itself.
Besides message $M_j$, $D_j$ comprises the protocol control information $L_j$ of the data-link layer for which integrity is demanded (also known as additional authenticated data), e.g., destination address, source address, and ethertype for MACsec, or the priority identifier and most of the control field (SDU Type, Virtual CAN Network ID, etc.) for CANsec, as well as the initialization vector $I_j$, which includes a freshness value (FV) to prevent replay attacks.

The origin authentication strategy we propose builds on a simplified version of TESLA that foresees immediate validation and delivery, as in \cite{2023-WFCS}, suitably enhanced to provide higher resistance to frame losses.
This permits it to be seamlessly layered atop MACsec and CANsec.
In particular, every frame $P_j$ additionally includes a key $K_j$ (i.e., $D_j = \langle L_j, I_j, K_j , M_j\rangle$) that, by applying a non-invertible hash function $H(\cdot)$, returns the key $K_{j-1}$ included in the previous frame $P_{j-1}$, i.e., $K_{j-1}=H(K_j)$. 
Since $H^{-1}(K_{j-1})$ can hardly be computed in the time interval between the transmission of two consecutive frames, if a message received on the \gls{sc} can be determined to be authentic (i.e., originated by $\Tsc$), then this property can be reliably verified for all subsequently received messages, provided that the underlying communication system ensures that they are strictly delivered in the original order.
Below we assume that key $K_0$, which is known as the \textit{root key}, is securely initialized by $\Tsc$ on all other members of the \gls{sc} (collectively denoted as $\Rsc$).
Conceptually, this corresponds to having a licit frame $P_{0}$ virtually delivered to all receivers, and permits them to easily assess the origin authenticity of every frame $P_{j}$ for ${j\geq 1}$.

Unlike TESLA, the above mechanism is extremely weak against man-in-the-middle (MITM) attacks: if some node in the network (e.g., a relay device like a switch) were able to temporarily delay a frame $P_j$ until $P_{j+1}$ has been sent, then it could forge (and propagate) a counterfeit frame $P^{\#}_{j}$ that includes key $K_{j}=H(K_{j+1})$ and looks licit.
However, above situation can not take place in small embedded systems where devices are interconnected through multidrop busses like CAN and 10Base-T1S Single Pair Ethernet (SPE) \cite{2023-Decker}, as those found onboard of vehicles in the automotive domain, since holding frames back in these cases is just impossible.
Although bridges can be possibly envisaged to interconnect a number of busses \cite{2023-WFCS-GC}, assuming that they can not be reprogrammed to behave as MITM once deployed is not unrealistic.
Moreover, we do not account for attacks that exploit physical access to the vehicle (like injection ones) to alter the network topology, e.g., by inserting a tampered relay device in the middle of a bus, as we are interested in remotely-delivered security threats.
Under these hypotheses, it is possible to safely neglect MITM attacks completely.

In the following, we will only focus on networked systems where one or more smart end-devices (sensors, cameras, ECUs, actuators, dashboards) have been compromised as a consequence of a remote attack, for instance exploiting their vulnerabilities and an active Internet connection, and can thus deliver in their turn masquerade attacks on the onboard network, which is assumed to work correctly.
As said before, these cases are expected to become more and more relevant in future scenarios that involve real automotive and embedded systems, whose operation is mostly stand-alone but that are nevertheless connected to the outside world.

A proper origin authentication strategy must meet a target security level, and this impacts on the size of the keys.
Generally speaking, increasing the key size improves security, but also implies higher bandwidth consumption and processing times.
Algorithms defined by the Secure Hash Standard (SHS) \cite{2015-SHS} are currently considered immune to pre-image attacks, but their digests take from $224$ to $\SI{512}{b}$.
Therefore, more space-efficient solutions are sought.
Concerning reliability, authentication mechanisms based on keychains like TESLA may cease working correctly due to repeated frame losses, which make the view of the keychain differ on the transmitter and receiver sides.
When this happens the \gls{sc} must be re-initialized, which implies that $\mathrm{T}_{\mathrm{SC}}$ must securely distribute the current root key to $\mathcal{R}_{\mathrm{SC}}$ again.
This procedure may take a non-negligible amount of time, during which transmission on the \gls{sc} is impaired and many messages may be dropped, negatively impacting on reliability.
Hence, every effort shall be undertaken to prevent this situation from occurring too often.

\section{Origin-Authentication Strategies}
\label{sec:Origin}
A number of significant simple multicast origin-authentication strategies derived from TESLA are described below that suit our design goals.
We start from the most basic approach and every strategy improves over the previous ones by addressing some specific weakness.
Sample diagrams illustrating the sequences of exchanged messages for the considered strategies are depicted in Fig.~\ref{fig:all}.

\subsection{Basic strategy}

\begin{figure}[]
    \centering
    \includegraphics[width=1.0 \columnwidth]{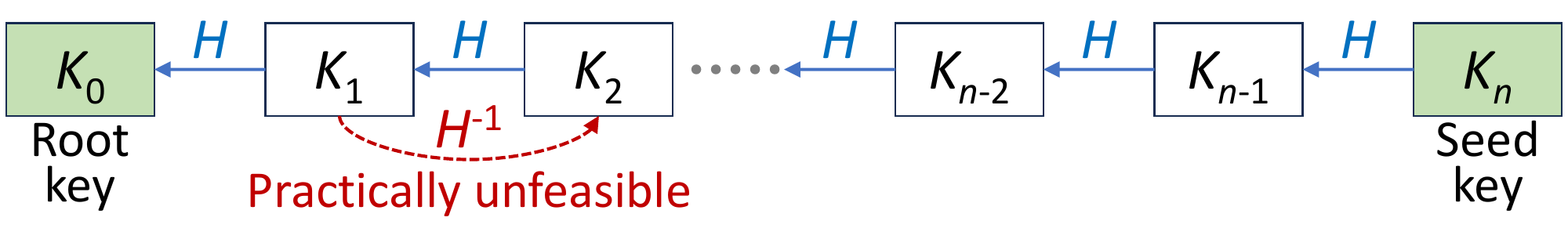}
    \caption{Keychain $\mathcal{C}$ for origin authentication in the SC.}
    \label{fig:chain}
\end{figure}

Let $\mathcal{C} = (K_i)_{i\in[0,n]}$ be a keychain $\Tsc$ generates from a random seed key $K_n$ (a nonce).
The \mbox{$i$-th} key in $\mathcal{C}$ is computed as $K_i = H^{n-i}(K_n)$ (see Fig. \ref{fig:chain}), where $H(\cdot)$ is a one-way (i.e., non-invertible) hash function like, e.g., SHA-256 \cite{2015-SHS}, and notation $H^m(k)$ indicates that the function is recursively applied  $m$ times to key $k$.
The last generated key, that is, $K_0 = H^n(K_n)$, is the root key, which is distributed in a secure way by $\Tsc$ to all the recipients in $\Rsc$ during the SC initialization phase (details of the initialization procedure are outside the scope of this paper).

In the most basic strategy, every \textit{origin-authenticated} frame (\mbox{A-frame}) includes a key taken from $\mathcal{C}$ in increasing $i$ order, which is encoded by $\Tsc$ in the part $D_j$ of the frame covered by the MAC, that is, $D_j = \langle L_j, I_j, i, K_i, M_j \rangle$.
Index $i$ permits receivers to know the position of $K_i$ in the keychain, e.g., $i=1$ means that the frame contains the key that immediately follows the root key, while $i=n$ indicates that the frame includes the last available key in $\mathcal{C}$.
Note that the authentication mechanism does not require receivers to know $n$ in order to work correctly (in theory, if a bound exists for the allowed values of $n$, not even $i$ is strictly required).

\subsubsection{Ideal case}
We consider initially the case when no frames are lost by the network.
Let us analyze what happens immediately after \gls{sc} initialization, that is, when $1 \leq j \leq n$. 
In this case, keys are drawn from $\mathcal{C}$ orderly, i.e., $i = j$ (for example, the initial frame includes $K_1$ as part of $D_1$).

Upon reception of frame $P_j$, its content is checked against the included MAC that relies on the shared key $K_\mathrm{SC}$ (this is accomplished by MACsec/CANsec).
Doing so verifies integrity of $D_j$ and, thanks to the freshness value, also excludes replay attacks.
In the following, we will assume that frames that failed these tests are discarded and not processed further.
However, receivers are still not guaranteed that the frame was sent by $\Tsc$, but only that the sender is part of the \gls{sc} (i.e., it knows $K_\mathrm{SC}$).
To verify the origin of the message a further check is needed on the receiver side, which involves the key $K$ taken from $D_j$ (not necessarily $K=K_i$, because of the possible presence of counterfeit messages).
Since we assume that no frames are lost, the test $H(K) = K_{i-1}$ can be performed ($K_{i-1}$ is the key included in the previous authentic frame).
If the test succeeds, the receiver is certain that frame $P_j$ comes from $\Tsc$, as we deem impossible for any node to compute $K_i = H^{-1}(K_{i-1})$ in the time elapsing between two consecutive frames.
This assumption can be often taken as true, because networked \glspl{ecu} are usually provided with limited computational resources, and can be verified since the design phase for cyclic message streams, whose periods typically range from few milliseconds to several seconds.
However, it is not unreasonable for sporadic streams as well: in fact, similarly to CANopen Process Data Objects, event timers can be defined for value-update messages that, upon expiry, force retransmission of the most recent available value.

This procedure is repeated for any message in the stream until $M_n$, when the end of keychain $\mathcal{C}$ is reached.
Then, a new keychain $\mathcal{C}'$ is generated by $\Tsc$, starting from a new nonce $K^{\prime}_{n}$, and the new root key $K'_0$ is distributed to $\Rsc$.
Unlike for \gls{sc} initialization, this can be done very easily by including $K'_0$ in frame $P_n$, along with $K_n$.
This way, receivers are guaranteed about the origin of $K'_0$.
Since frame $P_n$ securely connects a keychain to the next, we term it \textit{junction} frame (\mbox{J-frame}).
In this basic strategy, J-frames are characterized by having $j$ equal to an integer multiple  of $n$, that is, $j = l \cdot n$.
For them, $D_{l \cdot n} = \langle L_{l \cdot n}, I_{l \cdot n}, n, K_n, K'_0, M_{l \cdot n} \rangle$.

When the new keychain is effective (and before sending the first A-frame in the chain), index $i$ is reset to $1$ by $\Tsc$.
In the above strategy, the value of $i$ included with any message $M_j$ can be computed as $i=[(j-1) \bmod n] + 1$.
However, authentication strategies in this paper have been explicitly designed in such a way that receivers do not need to know $j$, but only $i$.
Moreover, although $i$ could be possibly derived from the freshness value used for managing the MAC, explicitly encoding its value in $D_j$ decouples origin authentication from MACsec/CANsec.
This allows these mechanisms to be kept separate and independent, and also increases robustness.

\begin{figure}[]
    \centering
    \includegraphics[width=1.0 \columnwidth]{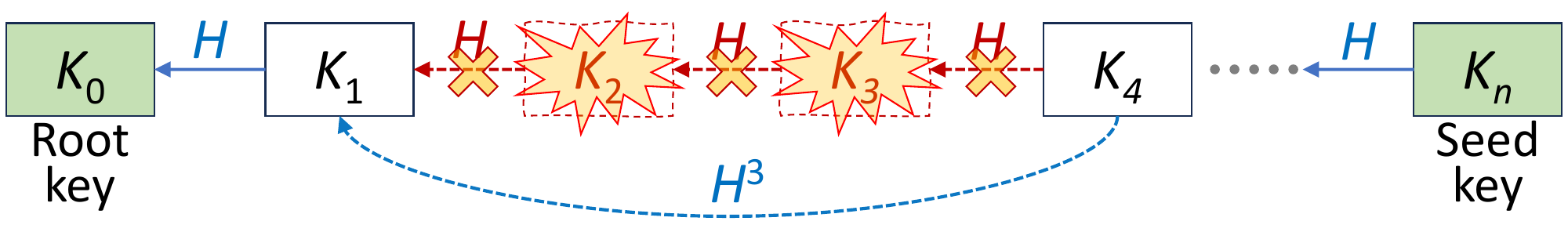}
    \caption{Robustness of keychain $\mathcal{C}$ against frame losses 
    (\textit{P}\textsubscript{2} and \textit{P}\textsubscript{3}).}
    \label{fig:chainerr}
\end{figure}

\begin{figure*}
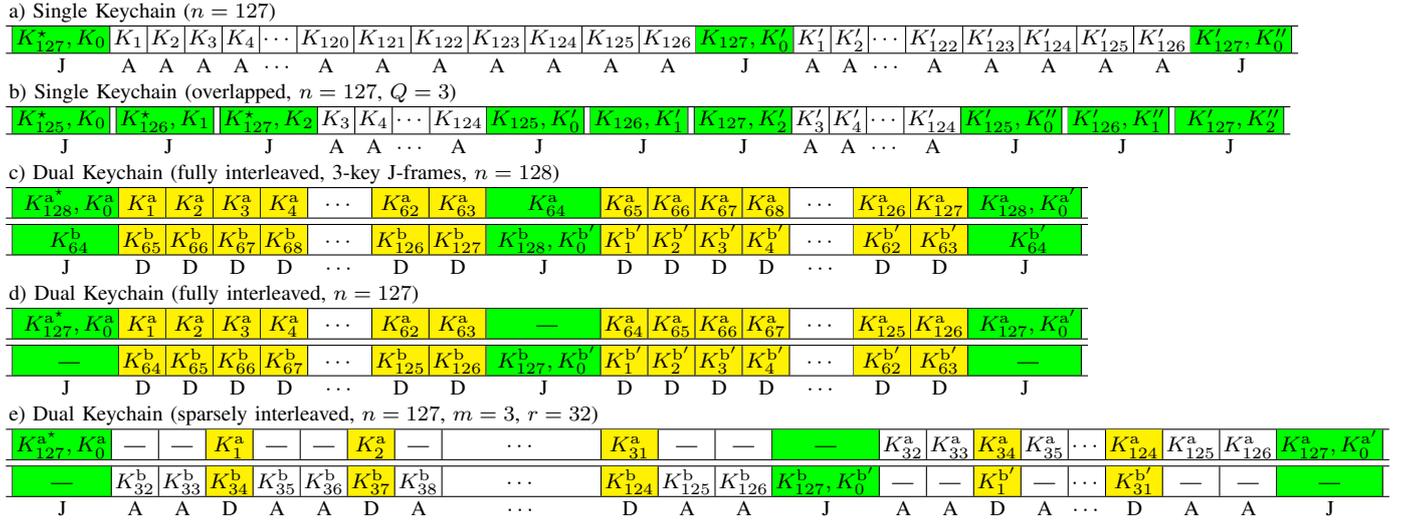

    \tabcolsep=0.04cm
    \renewcommand{\arraystretch}{1.1}    
    \footnotesize	

    \begin{tabular}{ c c c c c c c c c c c c c c c c c c c c c c c c c c c} 
    \multicolumn{10}{l}{a) Single Keychain ($n=127$)} \\      
    \hline
    \multicolumn{1}{c|}{} &  
    \multicolumn{1}{c|}{\cellcolor{green}$K^{\star}_{127},K_0$} &
    \multicolumn{1}{c|}{$K_{1}$} &  
    \multicolumn{1}{c|}{$K_{2}$} &  
    \multicolumn{1}{c|}{$K_{3}$} &  
    \multicolumn{1}{c|}{$K_{4}$} &  
    \multicolumn{1}{c|}{ $\cdots$ } &  
    \multicolumn{1}{c|}{$K_{120}$} &  
    \multicolumn{1}{c|}{$K_{121}$} &  
    \multicolumn{1}{c|}{$K_{122}$} &  
    \multicolumn{1}{c|}{$K_{123}$} &  
    \multicolumn{1}{c|}{$K_{124}$} &      
    \multicolumn{1}{c|}{$K_{125}$} &      
    \multicolumn{1}{c|}{$K_{126}$} &      
    \multicolumn{1}{c|}{\cellcolor{green}$K_{127},K'_0$}  &  
    \multicolumn{1}{c|}{$K'_{1}$} &
    \multicolumn{1}{c|}{$K'_{2}$} &  
    \multicolumn{1}{c|}{ $\cdots$ } &  
    \multicolumn{1}{c|}{$K'_{122}$} &      
    \multicolumn{1}{c|}{$K'_{123}$} &      
    \multicolumn{1}{c|}{$K'_{124}$} &      
    \multicolumn{1}{c|}{$K'_{125}$} &      
    \multicolumn{1}{c|}{$K'_{126}$} &      
    \multicolumn{1}{c|}{\cellcolor{green}$K'_{127},K''_0$} & \\
    \hline
    & J & A & A & A & A & $\cdots$ & A & A & A & A & 
      A & A & A & J & A & A & $\cdots$ & A & A & A & A & A & J \\
    \end{tabular}

    \begin{tabular}{ c c c c c c c c c c c c c c c c c c c c c c c c c c c c} 
    \multicolumn{10}{l}{b) Single Keychain (overlapped, $n=127$, $Q=3$) } \\      
    \hline
    \multicolumn{1}{c|}{} &  
    \multicolumn{1}{c||}{\cellcolor{green}$K^{\star}_{125},K_0$} &
    \multicolumn{1}{c||}{\cellcolor{green}$K^{\star}_{126},K_1$} &
    \multicolumn{1}{c|}{\cellcolor{green}$K^{\star}_{127},K_2$} &   
    \multicolumn{1}{c|}{$K_{3}$} &  
    \multicolumn{1}{c|}{$K_{4}$} &  
    \multicolumn{1}{c|}{ $\cdots$ } &  
    \multicolumn{1}{c|}{$K_{124}$} &      
    \multicolumn{1}{c||}{\cellcolor{green}$K_{125},K'_0$} &
    \multicolumn{1}{c||}{\cellcolor{green}$K_{126},K'_1$} &
    \multicolumn{1}{c|}{\cellcolor{green}$K_{127},K'_2$}  &  
    \multicolumn{1}{c|}{$K'_{3}$} &
    \multicolumn{1}{c|}{$K'_{4}$} &  
    \multicolumn{1}{c|}{ $\cdots$ } &  
    \multicolumn{1}{c|}{$K'_{124}$} &      
    \multicolumn{1}{c||}{\cellcolor{green}$K'_{125},K''_0$} & 
    \multicolumn{1}{c||}{\cellcolor{green}$K'_{126},K''_1$} & 
    \multicolumn{1}{c|}{\cellcolor{green} $K'_{127},K''_2$} & \\
    \hline
    & J & J & J & A & A & $\cdots$ & A &  
      J & J & J & A & A & $\cdots$ & A & J & J & J \\
    \end{tabular}

    \begin{tabular}{ c c c c c c c c c c c c c c c c c c c} 
    \multicolumn{10}{l}{c) Dual Keychain (fully interleaved, 3-key J-frames, $n=128$)} \\  
    \hline
    \multicolumn{1}{c|}{} &  
    \multicolumn{1}{c|}{\cellcolor{green} $K^{\mathrm{a}^\star}_{128}, K^{\mathrm{a}}_0$} &  
    \multicolumn{1}{c|}{\cellcolor{yellow}$K^{\mathrm{a}}_{1}$} & 
    \multicolumn{1}{c|}{\cellcolor{yellow}$K^{\mathrm{a}}_{2}$} & 
    \multicolumn{1}{c|}{\cellcolor{yellow}$K^{\mathrm{a}}_{3}$} &
    \multicolumn{1}{c|}{\cellcolor{yellow}$K^{\mathrm{a}}_{4}$} & 
    \multicolumn{1}{c|}{ $\cdots$ } & 
    \multicolumn{1}{c|}{\cellcolor{yellow}$K^{\mathrm{a}}_{62}$} &  
    \multicolumn{1}{c|}{\cellcolor{yellow}$K^{\mathrm{a}}_{63}$} &      
    \multicolumn{1}{c|}{\cellcolor{green} $K^{\mathrm{a}}_{64}$} & 
    \multicolumn{1}{c|}{\cellcolor{yellow}$K^{\mathrm{a}}_{65}$} & 
    \multicolumn{1}{c|}{\cellcolor{yellow}$K^{\mathrm{a}}_{66}$} & 
    \multicolumn{1}{c|}{\cellcolor{yellow}$K^{\mathrm{a}}_{67}$} & 
    \multicolumn{1}{c|}{\cellcolor{yellow}$K^{\mathrm{a}}_{68}$} & 
    \multicolumn{1}{c|}{ $\cdots$ } & 
    \multicolumn{1}{c|}{\cellcolor{yellow}$K^{\mathrm{a}}_{126}$} & 
    \multicolumn{1}{c|}{\cellcolor{yellow}$K^{\mathrm{a}}_{127}$} & 
    \multicolumn{1}{c|}{\cellcolor{green} $K^{\mathrm{a}}_{128}, K^{\mathrm{a}'}_0$} & \\
    \hline 
    \hline                        
    \multicolumn{1}{c|}{} &  
    \multicolumn{1}{c|}{\cellcolor{green} $K^{\mathrm{b}}_{64}$} & 
    \multicolumn{1}{c|}{\cellcolor{yellow}$K^{\mathrm{b}}_{65}$} & 
    \multicolumn{1}{c|}{\cellcolor{yellow}$K^{\mathrm{b}}_{66}$} & 
    \multicolumn{1}{c|}{\cellcolor{yellow}$K^{\mathrm{b}}_{67}$} & 
    \multicolumn{1}{c|}{\cellcolor{yellow}$K^{\mathrm{b}}_{68}$} & 
    \multicolumn{1}{c|}{ \; $\cdots$ \; } & 
    \multicolumn{1}{c|}{\cellcolor{yellow}$K^{\mathrm{b}}_{126}$} & 
    \multicolumn{1}{c|}{\cellcolor{yellow}$K^{\mathrm{b}}_{127}$} & 
    \multicolumn{1}{c|}{\cellcolor{green} $K^{\mathrm{b}}_{128}, K^{\mathrm{b}'}_0$} & 
    \multicolumn{1}{c|}{\cellcolor{yellow}$K^{\mathrm{b}'}_{1}$} & 
    \multicolumn{1}{c|}{\cellcolor{yellow}$K^{\mathrm{b}'}_{2}$} & 
    \multicolumn{1}{c|}{\cellcolor{yellow}$K^{\mathrm{b}'}_{3}$} & 
    \multicolumn{1}{c|}{\cellcolor{yellow}$K^{\mathrm{b}'}_{4}$} & 
    \multicolumn{1}{c|}{ \; $\cdots$ \; } & 
    \multicolumn{1}{c|}{\cellcolor{yellow}$K^{\mathrm{b}'}_{62}$} &     
    \multicolumn{1}{c|}{\cellcolor{yellow}$K^{\mathrm{b}'}_{63}$} &     
    \multicolumn{1}{c|}{\cellcolor{green} $K^{\mathrm{b}'}_{64}$} \\ 
    \hline
    & J & D & D & D & D & $\cdots$ & D & D & 
      J & D & D & D & D & $\cdots$ & D & D & J \\
    \end{tabular}

    \begin{tabular}{ c c c c c c c c c c c c c c c c c c c} 
    \multicolumn{10}{l}{d) Dual Keychain (fully interleaved, $n=127$)} \\  
    \hline
    \multicolumn{1}{c|}{} &  
    \multicolumn{1}{c|}{\cellcolor{green} $K^{\mathrm{a}^\star}_{127}, K^{\mathrm{a}}_0$} &  
    \multicolumn{1}{c|}{\cellcolor{yellow}$K^{\mathrm{a}}_{1}$} & 
    \multicolumn{1}{c|}{\cellcolor{yellow}$K^{\mathrm{a}}_{2}$} & 
    \multicolumn{1}{c|}{\cellcolor{yellow}$K^{\mathrm{a}}_{3}$} &
    \multicolumn{1}{c|}{\cellcolor{yellow}$K^{\mathrm{a}}_{4}$} & 
    \multicolumn{1}{c|}{ $\cdots$ } & 
    \multicolumn{1}{c|}{\cellcolor{yellow}$K^{\mathrm{a}}_{62}$} &  
    \multicolumn{1}{c|}{\cellcolor{yellow}$K^{\mathrm{a}}_{63}$} &      
    \multicolumn{1}{c|}{\cellcolor{green} ---} & 
    \multicolumn{1}{c|}{\cellcolor{yellow}$K^{\mathrm{a}}_{64}$} & 
    \multicolumn{1}{c|}{\cellcolor{yellow}$K^{\mathrm{a}}_{65}$} & 
    \multicolumn{1}{c|}{\cellcolor{yellow}$K^{\mathrm{a}}_{66}$} & 
    \multicolumn{1}{c|}{\cellcolor{yellow}$K^{\mathrm{a}}_{67}$} & 
    \multicolumn{1}{c|}{ $\cdots$ } & 
    \multicolumn{1}{c|}{\cellcolor{yellow}$K^{\mathrm{a}}_{125}$} & 
    \multicolumn{1}{c|}{\cellcolor{yellow}$K^{\mathrm{a}}_{126}$} & 
    \multicolumn{1}{c|}{\cellcolor{green} $K^{\mathrm{a}}_{127}, K^{\mathrm{a}'}_0$} & \\
    \hline 
    \hline                        
    \multicolumn{1}{c|}{} &  
    \multicolumn{1}{c|}{\cellcolor{green} ---} & 
    \multicolumn{1}{c|}{\cellcolor{yellow}$K^{\mathrm{b}}_{64}$} & 
    \multicolumn{1}{c|}{\cellcolor{yellow}$K^{\mathrm{b}}_{65}$} & 
    \multicolumn{1}{c|}{\cellcolor{yellow}$K^{\mathrm{b}}_{66}$} & 
    \multicolumn{1}{c|}{\cellcolor{yellow}$K^{\mathrm{b}}_{67}$} & 
    \multicolumn{1}{c|}{ \; $\cdots$ \; } & 
    \multicolumn{1}{c|}{\cellcolor{yellow}$K^{\mathrm{b}}_{125}$} & 
    \multicolumn{1}{c|}{\cellcolor{yellow}$K^{\mathrm{b}}_{126}$} & 
    \multicolumn{1}{c|}{\cellcolor{green} $K^{\mathrm{b}}_{127}, K^{\mathrm{b}'}_0$} & 
    \multicolumn{1}{c|}{\cellcolor{yellow}$K^{\mathrm{b}'}_{1}$} & 
    \multicolumn{1}{c|}{\cellcolor{yellow}$K^{\mathrm{b}'}_{2}$} & 
    \multicolumn{1}{c|}{\cellcolor{yellow}$K^{\mathrm{b}'}_{3}$} & 
    \multicolumn{1}{c|}{\cellcolor{yellow}$K^{\mathrm{b}'}_{4}$} & 
    \multicolumn{1}{c|}{ \; $\cdots$ \; } & 
    \multicolumn{1}{c|}{\cellcolor{yellow}$K^{\mathrm{b}'}_{62}$} &     
    \multicolumn{1}{c|}{\cellcolor{yellow}$K^{\mathrm{b}'}_{63}$} &     
    \multicolumn{1}{c|}{\cellcolor{green} ---} \\ 
    \hline
    & J & D & D & D & D & $\cdots$ & D & D & 
      J & D & D & D & D & $\cdots$ & D & D & J \\
    \end{tabular}                     

    \begin{tabular}{ c c c c c c c c c c c c c c c c c c c c c c c c c c c c c} 
    \multicolumn{10}{l}{e) Dual Keychain (sparsely interleaved, $n=127$, $m=3$, $r=32$)} \\      
    \hline
    \multicolumn{1}{c|}{} &  
    \multicolumn{1}{c|}{\cellcolor{green}$K^{\mathrm{a}^\star}_{127},K^{\mathrm{a}}_0$} &  
    \multicolumn{1}{c|}{---} & 
    \multicolumn{1}{c|}{---} & 
    \multicolumn{1}{c|}{\cellcolor{yellow}$K^{\mathrm{a}}_{1}$} &
    \multicolumn{1}{c|}{---} & 
    \multicolumn{1}{c|}{---} & 
    \multicolumn{1}{c|}{\cellcolor{yellow}$K^{\mathrm{a}}_{2}$} &  
    \multicolumn{1}{c|}{---} &     
    \multicolumn{1}{c|}{ $\cdots$ } &          
    \multicolumn{1}{c|}{\cellcolor{yellow}$K^{\mathrm{a}}_{31}$} &  
    \multicolumn{1}{c|}{---} &     
    \multicolumn{1}{c|}{---} &     
    \multicolumn{1}{c|}{\cellcolor{green}---} & 
    \multicolumn{1}{c|}{$K^{\mathrm{a}}_{32}$} & 
    \multicolumn{1}{c|}{$K^{\mathrm{a}}_{33}$} & 
    \multicolumn{1}{c|}{\cellcolor{yellow}$K^{\mathrm{a}}_{34}$} & 
    \multicolumn{1}{c|}{$K^{\mathrm{a}}_{35}$} & 
    \multicolumn{1}{c|}{ $\cdots$ } & 
    \multicolumn{1}{c|}{\cellcolor{yellow}$K^{\mathrm{a}}_{124}$} & 
    \multicolumn{1}{c|}{$K^{\mathrm{a}}_{125}$} & 
    \multicolumn{1}{c|}{$K^{\mathrm{a}}_{126}$} & 
    \multicolumn{1}{c|}{\cellcolor{green}$K^{\mathrm{a}}_{127},K^{\mathrm{a}'}_0$} & \\
    \hline 
    \hline                        
    \multicolumn{1}{c|}{} &  
    \multicolumn{1}{c|}{\cellcolor{green}---} & 
    \multicolumn{1}{c|}{$K^{\mathrm{b}}_{32}$} & 
    \multicolumn{1}{c|}{$K^{\mathrm{b}}_{33}$} & 
    \multicolumn{1}{c|}{\cellcolor{yellow}$K^{\mathrm{b}}_{34}$} & 
    \multicolumn{1}{c|}{$K^{\mathrm{b}}_{35}$} & 
    \multicolumn{1}{c|}{$K^{\mathrm{b}}_{36}$} & 
    \multicolumn{1}{c|}{\cellcolor{yellow}$K^{\mathrm{b}}_{37}$} & 
    \multicolumn{1}{c|}{$K^{\mathrm{b}}_{38}$} & 
    \multicolumn{1}{c|}{ $\cdots$ } & 
    \multicolumn{1}{c|}{\cellcolor{yellow}$K^{\mathrm{b}}_{124}$} & 
    \multicolumn{1}{c|}{$K^{\mathrm{b}}_{125}$} & 
    \multicolumn{1}{c|}{$K^{\mathrm{b}}_{126}$} & 
    \multicolumn{1}{c|}{\cellcolor{green}$K^{\mathrm{b}}_{127},K^{\mathrm{b}'}_0$} & 
    \multicolumn{1}{c|}{---} &     
    \multicolumn{1}{c|}{---} &     
    \multicolumn{1}{c|}{\cellcolor{yellow}$K^{\mathrm{b}'}_{1}$} &
    \multicolumn{1}{c|}{---} &     
    \multicolumn{1}{c|}{ $\cdots$ } &     
    \multicolumn{1}{c|}{\cellcolor{yellow}$K^{\mathrm{b}'}_{31}$} & 
    \multicolumn{1}{c|}{---} &     
    \multicolumn{1}{c|}{---} &           
    \multicolumn{1}{c|}{\cellcolor{green}---} \\ 
    \hline
    & J & A & A & D & A & A & D & A & $\cdots$ & D & A & A & 
      J & A & A & D & A & $\cdots$ & D & A & A & J \\
    \end{tabular}
    \caption{Comparison of data-origin authentication strategies 
        (only keys are reported, every column refers to a single A/D/J-frame).}
    \label{fig:all}    
\end{figure*}

\subsubsection{Lossy case}
Let us now consider the case when one or more frames may be lost by the network.
Frame transmissions in Ethernet are unconfirmed, which implies that losses go undetected.
A partial exception is found in CAN due to the error signaling mechanism (based on error frames), however: 
1) it can be disabled in CAN XL to make behavior similar to Ethernet, 
2) it does not work when CAN bridges are employed, and,
3) it does not cope with receive buffer overruns.
Whatever the case, efficient retransmission schemes can be hardly implemented for multicast transmissions.
In spite of this, the above authentication strategy proves to be quite robust against frame losses.
In fact, upon successful reception of frame $P_j$, the values of $i$ and $K$ are first extracted from $D_j$.
Then, check $H^{i}(K)=K_0$ is carried out: if they match the message origin is verified, otherwise the message is classified as non-authentic and discarded.

\subsection{Improving the basic strategy}
Repeatedly evaluating non-invertible hash functions like $H(\cdot)$ takes time.
Computation overhead can be lowered by letting every receiver store the values $i$ and $K$ of the last authentic received frame in two local variables $\hat{\iota}$ and $\hat{\kappa}$, respectively.
For J-frames, $\hat{\iota}=0$ and $\hat{\kappa}$ is set to the newly generated root key $K'_0$. 
By exploiting such information, backtracking is performed on frame arrival, which corresponds to checking whether
$H^{i-\hat{\iota}}(K)=\hat{\kappa}$.
An example is shown in Fig.~\ref{fig:chainerr}: upon $P_1$ reception, local variables are set to $\hat\iota=1$ and $\hat\kappa=K_1$.
If $P_2$ and $P_3$ are lost in a row, when $P_4$ is received function $H$ is applied $4-1=3$ times to $K_4$ and the result is checked against the latest available key $\hat\kappa$ (that corresponds to $K_1$).
Since backtracking can not get past the root key $K_0$ of the current keychain, the test shall fail any time $i \leq \hat{\iota}$.
In fact, every frame that cannot be authenticated must be treated as non-authentic.

As a further optimization to avoid wasting time in the case of severe losses, a counter $c$ encoded on $w_c$ bits (e.g., one octet, can be included in $D_j$, which $\Tsc$ increases by one (modulo $2^{w_c}$) after sending a J-frame (that is, every time a new keychain is started).
On the recipients' side, whenever a \mbox{frame} is received that passes origin authentication, a local state variable $\hat c$ is updated.
For A-frames $\hat c$ is set equal to $c$, whereas for \mbox{J-frames} it is set to $(c+1) \bmod 2^{w_c}$.
Exploiting such information, authentication fails for sure if $c \neq \hat{c}$, as happens when one or more \mbox{J-frames} are lost.
If the number of dropped J-frames is so high that $c$ overflows and becomes equal to $\hat{c}$ again (undermining above test), backtracking on $H(\cdot)$ still allows to determine that the frame is non-authentic, safeguarding security.
According to this approach, which prevents multiple calls to $H(\cdot)$ when unnecessary, the validation function is defined as
\begin{align}
    \label{eq:testsingle}
    V(K) \doteq & \left[ (c = \hat{c}) \wedge (i > \hat{\iota}) \wedge  (H^{i-\hat{\iota}}(K) = \hat{\kappa}) \right],
\end{align}
whose evaluation takes place left-to-right and stops as soon as the result is false.
From now on, we will assume that $c$ is included in $D$, i.e., $D_j = \langle L_j, I_j, c, i, K_i, M_j \rangle$.
It is worth stressing that, as variables $\hat{c}$, $\hat{\iota}$, and $\hat{\kappa}$ are only updated on authentic frames, attacks aimed at poisoning the local keychain status in receivers are prevented.

Fig.~\ref{fig:all}.a refers to the basic strategy with a single keychain where $n=127$.
Below any frame (corresponding to a column of the diagram), its type (A or J) is specified.
Depending on the type, one (A) or two (J) keys are included in the frame.
Superscripts `$^\star$' and `$''$' denote the keychain before $\mathcal{C}$ and the one following $\mathcal{C}'$, respectively.

\subsection{Preserving keychain status coherence}
As can be easily seen, the above strategy is unable to tackle situations where even a single \mbox{J-frame} is lost.
In fact, in the absence of it the new keychain is not started in receivers, which means that all subsequent A-frames (and J-frames as well) will be discarded.
To recover communication (making the system operational again), the origin authentication mechanism for the \gls{sc} must be re-initialized, which may take some time.

To detect above situation, several techniques can be used by receivers in $\Rsc$. 
The simplest one is to check if no valid frames are received for a while.
In case transmissions from $\Tsc$ are periodic, a timeout can be set every time a valid frame is received, whose duration $T_\mathrm{to}$ is larger than the period $T$.
Heuristics could be possibly devised to determine $T_\mathrm{to}$ based on the number  of elements left in the current keychain, e.g., $T_\mathrm{to} = (n-i+1)T$, which corresponds to the worst-case arrival time of the next \mbox{J-frame}.
In fact, invoking re-initialization before $\mathcal{C}$ ends is pointless.
Upon timeout expiry, the involved receivers will ask $\Tsc$ to securely provide them an up-to-date keychain status, e.g., a certified tuple $\langle c',i',K'_i \rangle$ that reflects its current view about $\mathcal{C}'$.
However, in case of sporadic transmissions, this approach is likely to often cause \gls{sc} recovery to occur uselessly (if $T_\mathrm{to}$ is chosen too short), which should be avoided, or late (if $T_\mathrm{to}$ is too long), which may cause several potentially valid \mbox{A-frames} (belonging to $\mathcal{C}'$, $\mathcal{C''}$, etc.) to be dropped, negatively impacting on reliability.

To shorten reaction times, re-initialization can be invoked when a frame is received for which condition \eqref{eq:testsingle} is false, which denotes either the misalignment of the keychains at the transmitting and receiving sides or the arrival of a counterfeit message.
Although \gls{sc} recovery is performed as soon as it is (seemingly) required (i.e., when  validation fails), attackers in $\Rsc$ may exploit this mechanism by repeatedly sending fake messages which, even if discarded by the other receivers, nevertheless enforce them to re-initialize the \gls{sc}.
This enables DoS attacks that cannot be easily detected by an intrusion detection system (IDS), e.g., by analyzing the rate at which messages are seen on the bus.

A reasonable trade-off is to have receivers in $\Rsc$ set a short timeout (e.g., $T_\mathrm{to}=\overline{T}$, where $\overline{T}$ is the mean iterarrival time) when condition \eqref{eq:testsingle} is false, and only after its expiry ask for re-initialization.
If a valid frame is received in the meanwhile, the timer is stopped and the keychain is retained, since this implies that  \eqref{eq:testsingle} failed due to a counterfeit message.
This approach has almost the same advantages as the previous one (just $T_\mathrm{to}$ slower), but is mostly immune to above DoS attacks. 
It must be noted that above approaches are not meant to deal with message losses at the application level, but only to promptly recover from situations where the \gls{sc} gets broken.

\subsection{Overlapping keychains}
A simple way to decrease the likelihood of issues due to \mbox{J-frame} losses in single keychain strategies is to foresee a number $Q > 1$ of such frames, which are sent on the \gls{sc} with no \mbox{A-frames} in between, enabling receivers to switch keychain even in the case some of them are dropped.
This corresponds to make adjacent keychains overlap by $Q$ elements ($Q=1$ corresponds to the basic strategy).
In particular, when considering $\mathcal{C}$ and $\mathcal{C}'$, the root key $K'_0$ of $\mathcal{C}'$ is included in a J-frame along with key $K_{n-Q+1}$ of $\mathcal{C}$, and the same holds for the next $Q-1$ keys (i.e., keys $K'_{i}$ and $K_{n-Q+1+i}$ are paired for $0\leq i \leq Q-1$).
Note that a simpler solution can be devised where $K'_0$ is included in every J-frame of the entire overlapped portion.
Doing so is not recommended, as the time available to attackers to compute $K'_1 = H^{-1}(K'_0)$ increases up to $Q \cdot T$.

In Fig.~\ref{fig:all}.b an example of single overlapped keychain strategy is shown, where $n=127$ and $Q=3$.
When this kind of strategy is adopted, synchronization is lost only if all $Q$ adjacent J-frames are dropped, which is typically unlikely to happen if $Q$ is chosen suitably.
Nevertheless, in the presence of error bursts this event cannot be ruled out.

\begin{figure*}[]
    \centering
    \includegraphics[width=1.0 \textwidth]{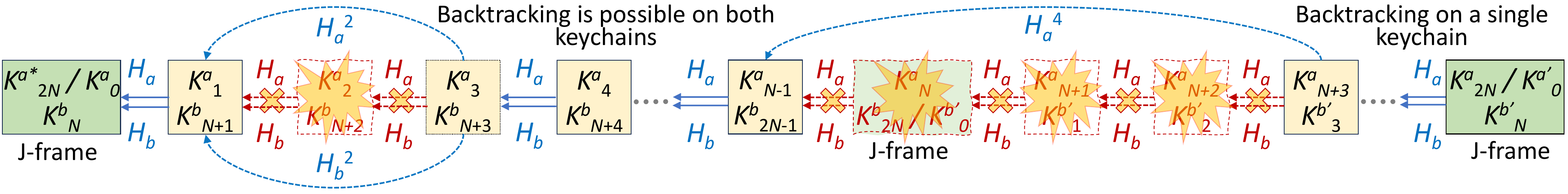}
    \caption{Robustness of dual keychain against frame losses: backtracking can be available on either two keychains (left) or a single one (right).}
    \label{fig:chaindouble}
\end{figure*}

\subsection{Exploiting a dual keychain}
To further improve robustness against frame losses, a dual keychain strategy can be used which relies on two keychains $\mathcal{C}_\mathrm{a}$ and $\mathcal{C}_\mathrm{b}$.
The length of each keychain is set to $n=2N$, so that the number of included keys is even.
Keychains are displaced by $N$ elements, thus providing circular symmetry.
This can be seen in the example of Fig.~\ref{fig:all}.c, where $n=128$.

Any \textit{dual authenticated} frame (D-frame) includes a pair of keys $K^\mathrm{a}$ and $K^\mathrm{b}$.
In principle, it should also include two pairs $\langle c_\mathrm{a}, i_\mathrm{a} \rangle$ and $\langle c_\mathrm{b}, i_\mathrm{b}  \rangle$ but, by exploiting symmetry, only values $i_\mathrm{a}$ and $c_\mathrm{a}$ need to be encoded in the frame (we denote them $i$ and $c$, respectively), whereas $i_\mathrm{b}$ and $c_\mathrm{b}$ can be easily derived.
Referring again to the case $1 \leq j \leq 2N$ (where $i=j$), the sequence of frames can be split in two halves (see Fig.~\ref{fig:chaindouble}).
In the \textit{first half} ($1 \leq i \leq N$), $i_\mathrm{b} = i_\mathrm{a} + N$, $c_\mathrm{b} = c_\mathrm{a}$, and $D_j = \langle L_j, I_j, c, i, K^\mathrm{a}_i, K^\mathrm{b}_{i+N}, M_j \rangle$.
Instead, in the \textit{second half} ($N+1 \leq i \leq 2N$), $i_\mathrm{b} = i_\mathrm{a} - N$, $c_\mathrm{b} = c_\mathrm{a} + 1$, and $D_j = \langle L_j, I_j, c, i, K^\mathrm{a}_i, K^{\mathrm{b}'}_{i-N}, M_j \rangle$.

This strategy foresees two distinct J-frames, one for each keychain.
A new keychain $\mathcal{C}'_{\mathrm{a}}$ is started when $i=2N$, at which time $c$ is increased by one.
Instead, the transition from $\mathcal{C}_\mathrm{b}$ to $\mathcal{C}'_{\mathrm{b}}$ takes place when $i=N$, at the border between the first and the second half.
Possibly, two distinct hash functions $H_\mathrm{a}(\cdot)$ and $H_\mathrm{b}(\cdot)$ can be used to generate and verify keys belonging to the two keychains.
This could bring slight benefits to security, but it also makes implementations more complex.
Therefore, in the following the same hash function $H(\cdot)$ is considered for both keychains.

As depicted in Fig.~\ref{fig:chaindouble}, the strategy works as long as authentication succeeds on one of the two keychains.
If a sequence of consecutive frames is lost, that does not include any \mbox{J-frames} (see the left part of the figure), backtracking is possible on both keychains.
Conversely, if a single J-frame is lost (on the right of the figure), backtracking must be done on the  unaffected keychain  ($\mathcal{C}_\mathrm{a}$ in the example).
This implies that \gls{sc} recovery is required only when two (or more) subsequent \mbox{J-frames} and all intermediate \mbox{D-frames} are lost, i.e., all error bursts affecting up to $N$ adjacent frames are tolerated.
Performing validation on two keys at the same time (using both $\mathcal{C}_\mathrm{a}$ and $\mathcal{C}_\mathrm{b}$) brings negligible improvements to security with respect to the case when a single key is considered.
In fact, if a brute force attack is carried out, guessing two $\SI{256}{b}$ keys requires on average the same time as guessing a single $\SI{257}{b}$ key (and drastically less time than guessing a single $\SI{512}{b}$ key).

The main disadvantage of the dual-keychain strategy is that, J-frames include three keys. 
This increases both overheads and transmission duration, impacting, e.g., on feasibility analysis when dealing with  real-time applications.
An alternative approach is to include in J-frames only the two keys of the keychain being joined.
Referring to the J-frame determining the transition of $\mathcal{C}_\mathrm{b}$ in the middle of Fig.~\ref{fig:chaindouble}, this means including just $K^\mathrm{b}_{n}$ and $K^{\mathrm{b}'}_0$ (in this case $n=2N-1$).

Not having the two keychains guarding each other in \mbox{J-frames} has no practical effects on security.
Also reliability is marginally affected since, again, two subsequent J-frames and all the intermediate D-frames have to be lost in order to break both keychains and enforce \gls{sc} recovery.
Hence, in the following we will refer to this second variant of the fully-interleaved dual-keychain strategy, an example of which is shown in Fig.~\ref{fig:all}.d with $n=127$.
As can be seen, error bursts that cause up to $N-1$ frames to be dropped ($63$, in the example) are tolerated, bringing a tangible improvement over the single overlapped keychain strategy.

Since backtracking must be supported on both keychains, two tuples of local variables are needed on receivers to keep their state, namely, $\langle \hat{c}_\mathrm{a}, \hat\iota_\mathrm{a}, \hat\kappa_\mathrm{a} \rangle$ and $\langle \hat{c}_\mathrm{b}, \hat\iota_\mathrm{b}, \hat\kappa_\mathrm{b} \rangle$.
Similarly to the basic strategy, when a \mbox{J-frame} is received and after it has been successfully authenticated, the state of the keychain $\mathcal{C}_x$ being joined, where $x\in\{a,b\}$, is updated to $\langle \hat{c}_x= (c_x+1) \bmod 2^{w_c}, \hat\iota_x=0, \hat\kappa_x = K'_0 \rangle$.
Instead, the state of the other keychain is left untouched, as no relevant information is available in the J-frame for it.
On \gls{sc} initialization, the root keys for $\mathcal{C}_\mathrm{a}$ and $\mathcal{C}_\mathrm{b}$ are set to $K^\mathrm{a}_0$ and $K^\mathrm{b}_{N-1}$, respectively.
Frame authentication only needs to succeed on a single keychain, and the validation function becomes
\begin{align}
    \label{eq:testdual}
    & V(K^\mathrm{a},K^\mathrm{b}) \doteq 
    \left[ (c_\mathrm{a} = \hat{c}_\mathrm{a}) \wedge (i_\mathrm{a} > \hat{\iota}_\mathrm{a}) \wedge \left( H^{i_\mathrm{a}-\hat{\iota}_\mathrm{a}}(K^\mathrm{a}) = \hat{\kappa}_\mathrm{a} \right) \right]
    \nonumber \\ 
    & \;\;\;\;\;\;  \vee \left[ (c_\mathrm{b} = \hat{c}_\mathrm{b}) \wedge (i_\mathrm{b} > \hat{\iota}_\mathrm{b}) \wedge \left( H^{i_\mathrm{b}-\hat{\iota}_\mathrm{b}}(K^\mathrm{b}) = \hat{\kappa}_\mathrm{b} \right) \right]. 
\end{align}

The order with which terms are evaluated in the above logical disjunction is in theory irrelevant.
Practically, it should be selected in such a way to minimize backtracking, that is, the number $i-\hat{\iota}$ of calls to the hash function $H (\cdot)$.

A simpler solution is in theory possible that derives from the basic strategy (single keychain $\mathcal{C}$), where every \mbox{A-frame} is replaced by a \mbox{J-like frame} that also bears the root key $K'_0$ of the next keychain $\mathcal{C}'$ (commitment to $K'_1$ is repeated $n$ times before $K'_1$ is disclosed).
However, doing so worsens the security level and should be avoided, as the attacker is given much more time (up to $n \cdot T$) to try evaluating $K'_1 = H^{-1}(K'_0)$.

\subsection{Reducing Overheads}
The above dual-keychain strategy features very good robustness, but it is quite expensive in terms of both communication and computational resources.
In fact, keys are quite long ($\SI{32}{B}$ for SHA-256) and a pair of them is always included in every J/D-frame.
Besides consuming bandwidth, when the bit rate is set to $\SI{10}{Mb/s}$ the duration of frame transmissions increases by about $\SI{25.6}{\mu s}$ with respect to the single keychain case.

A compromise solution can be devised that reduces overheads sensibly at the price of a slightly worse capability to recover from frame losses.
The basic idea is to sparsely interleave the keys of $\mathcal{C}_\mathrm{a}$ and $\mathcal{C}_\mathrm{b}$, in such a way that only sometimes two keys are included in the same frame.
Several heuristics are possible to this aim, characterized by different \mbox{A/D/J-frame} patterns.
In Fig.~\ref{fig:all}.e an example strategy is shown that offers specific advantages, although other design choices are possible as well.
The frame pattern on the \gls{sc} is made up of basic sequences, each one including $m$ frames.
At the beginning of every sequence there are $m-1$ \mbox{A-frames}, while the last one is a \mbox{D-frame} that permits to cross-validate the two keychains.
This basic sequence is repeated $r$ times and, in the last repetition, the D-frame is replaced by a J-frame.

As can be seen in the figure, in the first half of the keychain every frame includes a key from $\mathcal{C}_\mathrm{b}$, whereas those from $\mathcal{C}_\mathrm{a}$ are included only in \mbox{D-frames}.
The situation is reversed in the second half.
The reason for this arrangement is that, the first half immediately follows a J-frame aimed at joining $\mathcal{C}_\mathrm{a}$.
If this J-frame is lost, all the following A-frames that use $\mathcal{C}_\mathrm{a}$ are set as non-authentic and discarded (up to the reception of a valid \mbox{D-frame}).
Conversely, at that time $\mathcal{C}_\mathrm{b}$ has a high probability to be correctly set in receivers, since in the recent past it was given many opportunities to re-synchronize through \mbox{J/D-frames}.
Making its keys available for validation in every frame makes false negatives for $V(\cdot)$ unlikely, hence improving the reliability level.
All \mbox{J/D-frames} in the first half (we denote the start-up phase for $\mathcal{C}_\mathrm{a}$) have to be regarded as a robust way to initialize that keychain.
In this way, $\mathcal{C}_\mathrm{a}$ will be (with high probability) correctly synchronized by the time the next \mbox{J-frame} (that joins $\mathcal{C}_\mathrm{b}$) arrives.
In the second half (start-up phase for $\mathcal{C}_\mathrm{b}$), the above reasoning still holds by inverting the roles of $\mathcal{C}_\mathrm{a}$ and $\mathcal{C}_\mathrm{b}$.
The above strategy has a drawback, due to the larger time between commitment and disclosure in the start-up part of the chain, where keys are sent sparsely.
This makes it worth not increasing $m$ excessively.

\subsection{Key Usage Efficiency}
One of the metrics to evaluate above authentication strategies is the \textit{key transmission efficiency} $\eta_\mathrm{KT}$, defined as the inverse of the mean number of keys that are sent per message ($\eta_\mathrm{KT}=1$ is obtained for an infinitely long single keychain).
Index $\eta_\mathrm{KT}$ provides a measure of the average bandwidth consumption due to origin authentication.
Let $n$ be the length of the keychain excluding the root key.
For the single keychain strategy in Fig.~\ref{fig:all}.a, one J-frame is required every time the keychain is restarted, therefore $\eta_\mathrm{KT} = n/(n+1)$.
Higher $n$ implies better efficiency.
For example, if $n=127$ then $\eta_\mathrm{KT}=0.99219$, whereas $\eta_\mathrm{KT}=0.99609$ when $n=255$. 
However, since $\Tsc$ must compute and store the entire keychain in advance, $n$ directly impacts on memory consumption of this node (computation time is a less critical aspect, since the keychain can be created incrementally by a background process in parallel to communication). 

When the single overlapped keychain in Fig.~\ref{fig:all}.b is considered, efficiency decreases and $\eta_\mathrm{KT} = (n-Q+1)/(n+1)$.
For example, if $n=127$ and $Q=3$ then $\eta_\mathrm{KT}=0.97656$, while increasing $Q$ to $16$ makes efficiency decrease to $\eta_\mathrm{KT}=0.875$.

In the fully interleaved dual keychain in Fig.~\ref{fig:all}.d every frame (either J or D) includes exactly two keys, hence $\eta_\mathrm{KT} = 0.5$. 
Finally, for the sparsely interleaved dual keychain in Fig.~\ref{fig:all}.e the fraction of J/D-frames (that include two keys) is equal to $1/m$.
As a consequence, $\eta_\mathrm{KT}=m/(m+1)$.
For example, when $m=3$ (as in Fig.~\ref{fig:all}.e, where $32$ synchronization opportunities are provided to each keychain, counting the initial \mbox{J-frame} and the following $31$ D-frames) then $\eta_\mathrm{KT}=0.75$.
Increasing $m$ to $7$ raises efficiency to $\eta_\mathrm{KT}=0.875$, which is exactly the same as using a single overlapped keychain (where \mbox{J-frames} are contiguous) with $Q=16$.
The main advantage of the sparse dual-interleaved keychain is that, it is noticeably more robust against error bursts, since J/D-frames are not sent back-to-back, and better supports never-give-up strategies.

\subsection{Security analysis}
So far, we have mostly considered an attacker who want to discover the next key.
This permits successfully injecting a single fake message.
However, the most effective way to trick origin authentication strategies is to look for the seed key $K_n$ of the current keychain given its root key $K_0$.
Let $|K|$ refer to the key size (e.g., $128$ for MD5).
Since the seed is a random integer in the range $\left[ 0,2^{|K|}-1 \right]$, dictionary attacks are useless.
Instead, a brute-force pre-image attack is required, as described by the pseudo-code in Fig.~\ref{fig:pseudo-code-att}.
Every time a \mbox{J-frame} is received, the attacker generates a keychain starting from a random key, until either $K_0$ is found or another \mbox{J-frame} arrives.
Generated keys are orderly added to a circular buffer with $n$ elements (overwriting the existing content), making the last $n$ keys available upon success (\texttt{return} instruction), which correspond to the desired keychain.
In particular, the oldest element in the buffer is the seed $K_n$, which permits the attacker to forge a valid \mbox{J-frame} and take control of the \gls{sc}.

The time available to compromise a keychain corresponds to its lifetime $T_\mathrm{C}$.
For the basic strategy $T_\mathrm{C} = n \cdot \overline{T}$, while for the others it is slightly longer.
Let $R_\mathrm{H}$ denote the processing speed for $H(\cdot)$, expressed in hashes per second (H/s).
In practice, $R_\mathrm{H}$ may range from less than $\SI{1}{MH/s}$ to many $\SI{}{TH/s}$, depending on whether above code is executed on the embedded system (e.g., a car's ECU) or outside it (e.g., by exploiting an active Internet connection), and, in the latter case, on parallelization (using GPUs and, possibly, multiple cracking rigs).

The probability for a single random attempt to guess a key is $P_\mathrm{S} = 2^{-|K|}$, which is extremely small.
While a keychain is in use, $N_\mathrm{H} = R_\mathrm{H} \cdot T_\mathrm{C}$ such attempts can be done to compromise it.
The probability that at least one of them succeeds to produce $K_0$ (after computing at least $n$ keys) can be approximated as $P_\mathrm{C} \simeq N_\mathrm{H} \cdot P_\mathrm{S}$.
Similarly, the rate with which keychains can be compromised (kind of a failure rate for authentication) is
\begin{align}
    \lambda_\mathrm{C} = \frac{P_\mathrm{C}}{T_\mathrm{C}} = \frac{R_\mathrm{H}}{2^{|K|}},
\end{align}
and $1/\lambda_\mathrm{C}$ is mean time between failures (MTBF), which corresponds to the average time before authentication is compromised.
If MD5 is considered, and assuming $R_\mathrm{H} = \SI{1000}{TH/s}$, the MTBF is $1.079\cdot 10^{16}$ years, which implies that there is seemingly no need to move to solutions with larger keys like SHA-2.
The option to use the more compact SHAKE128 \cite{2015-SHA3} with $\SI{96}{b}$ digest length, which in above conditions ensures, in theory, a $2.5$ million year MTBF, could be also investigated.
However, this is out of the scope of this paper.

\begin{figure}[]
    \footnotesize
\begin{alltt}
while(1) // code below can be parallelized
  k:=random(\(0...2\sp{|K|}-1\)) // random seed key
  repeat               // try finding the root
    append(k)            // save keys of keychain
    k:=H(k)            // move toward the root
    if(k=K\(\sb0\)) return    // attack succeeded!
  until(next K\(\sb0\) is available) // next J-frame
\end{alltt}
    \caption{Pseudo-code for the brute-force pre-image attack.}
    \label{fig:pseudo-code-att}
\end{figure}

\section{Unified Architecture}
\label{sec:Unified}
Authentication strategies described above and reported in Fig.~\ref{fig:all} provide different trade-offs between complexity and reliability.
As can be seen, in all cases the adopted strategy is decided completely by the transmitter $\Tsc$.
It is interesting to show that a \textit{unified} receiver can be devised that suits all strategies at the same time. 
This aspect is quite appealing, as it leaves $\Tsc$ the option to change  the authentication strategy at runtime depending, e.g., on the actual (measured) network conditions (for example, the perceived failure rate), hence featuring adaptive behavior.
Details about the TRUDI architecture and operation are provided below.

The proposed strategy supports an arbitrary number of keychains, and all the previous kinds of frames (A, D, J, etc.) are replaced by a single \textit{unified authentication} frame (\mbox{U-frame}).
Let $G$ be the maximum number of keychains that can be used contemporarily on the \gls{sc}, defined when it is initialized.
An integer $g \in [1,G]$ can be used to uniquely identify a keychain $\mathcal{C}_{g}$.
Such identifiers are temporary, that is, they can be reassigned by $\Tsc$ at runtime, in such a way to limit the amount of resources to be allocated on receivers.
One may argue that reusing identifiers may lead to discrepancies between the transmitting and receiving sides as a consequence of frame losses.
This is not an issue, since backtracking performed using $H$ always prevents situations where security is compromised: at worst, \gls{sc} recovery is performed.

Let $A_{j}^{g} = \langle \tau_{j,g}, \Omega_{j,g}; c_{j,g}, i_{j,g}, K_j^{g} \rangle$ be the information included by $\Tsc$ in frame $P_j$ that concerns keychain $\mathcal{C}_{g}$.
The \textit{TX status} flag~$\tau_{j,g}$ is a Boolean value that specifies whether or not information in $A_{j}^{g}$ is significant.
Clearly, more space-effective representations will be used in practice, which consume as little space as possible in the frame when $\tau_{j,g}$ is false.
The \textit{termination} flag~$\Omega_{j,g}$ is also Boolean: if true, no additional information will be sent by $\Tsc$ about $\mathcal{C}_{g}$ after $A_{j}^{g}$ (for example, because the keychain is finished).
This flag is used to force receivers to dismiss the keychain, invalidating its local state and freeing identifier $g$ (that can be reused).
This permits, e.g., to distinguish between J-like and D-like frames: in both cases the U-frame includes two keys, but in the former the oldest keychain is set for deletion.
Set $\mathcal{A}_j = \{ A_j^{g}, g \in [1,G] \}$ includes all the information needed for origin authentication related to every active keychain that is included in a \mbox{U-frame}.
Then, $D_j = \langle L_j, I_j, \mathcal{A}_j, M_j \rangle$.

Every receiver maintains the state of all keychains in use by means of local variables.
In particular, the state of keychain $\mathcal{C}_{g}$ is described by the tuple $S^{g} = \langle \varrho_g; \hat{c}_g, \hat\iota_g, \hat\kappa_{g} \rangle$, whose elements have similar meanings as previously described status variables.
The \textit{RX status} flag $\varrho_g$ is a Boolean that specifies whether or not the information stored in $S^{g}$ is significant.
Only when $\varrho_g$ is true the related keychain can be used for authentication.
On \gls{sc} initialization, the state of one or more keychains are suitably set by $\Tsc$ (the case when no keychain is initialized is pointless, as no prior information is made available for backtracking).
For keychains that were not initialized or that have been dismissed, $\varrho_g$ is set to false.

Whenever a \mbox{U-frame} is received (and after MAC validation) set $\mathcal{A} = \{ A^{g}, g \in [1,G] \}$, where $A^{g} = \langle \tau_{g}, \Omega_{g}; c_{g}, i_{g}, K^{g} \rangle$, is extracted from $D$ (the receiver does not need to know $j$) and its origin is authenticated by calling a validation function $V$ that iteratively considers all the keys included in the frame 
\begin{align}
    \label{eq:testuniv}
    V(\mathcal{A}) \doteq &
    \bigvee_{g \in [1,G]} \left[\tau_g \wedge \varrho_g \wedge 
    V_\circ \left( A^{g} \right) \right],
\end{align}
where $V_\circ$ is used to check any single key, i.e.,
\begin{align}
    \label{eq:testone}
    V_\circ \left( A^{g} \right) \doteq   
    (c_g = \hat{c}_g) \wedge (i_g > \hat{\iota}_g) \wedge \left( H^{i_g-\hat\iota_g} (K^{g}) = \hat\kappa_g \right). \nonumber  
\end{align}
Evaluation terminates as soon as a term of logical disjunction in \eqref{eq:testuniv} is found that evaluates to true.
As said before, the order with which keys are validated is irrelevant for authentication, but may impact on computation time.

\begin{figure}[]
    \footnotesize
\begin{alltt}
while(1)
  P:=recv() // a new frame is received
  <L,I,A,M,MAC>:=decode(P)       // decode frame
  if(!validate(L,I,A,M,MAC,K\(\sb\mathrm{SC}\))) // check integrity
    continue      // discard invalid frame
  v:=FALSE        // initially assume non-authentic
  for(g:=1 to G)  // check origin authenticity
    if(A[g].\(\tau\,\wedge\,\)S[g].\(\varrho\,\wedge\,\)A[g].c=S[g].\(\hat{c}\,\wedge\,\)A[g].i>S[g].\(\hat{\iota}\))
      k:=A[g].K
      for(l:=1 to A[g].i-S[g].\(\hat\iota\))  // compute H\(\sp\mathit{n}\)
        k=H(k) 
      if(k=S[g].\(\hat\kappa\))
        v:=TRUE   // one valid key is enough
        break
  if(v)  // retain authentic frame
    reset_timeout()
    for(g:=1 to G)  // update local keychain state
      if(A[g].\(\tau\))
        if(A[g].\(\Omega\))  // dismiss keychain
          S[g].\(\varrho\):=FALSE
        else
          S[g].\(\varrho\):=TRUE
          S[g].\(\hat{c}\):=A[g].c
          S[g].\(\hat{\iota}\):=A[g].i
          S[g].\(\hat\kappa\):=A[g].K
    deliver(M) // pass message to upper layers
  else    // discard non-authentic frame
    if(!is_timeout_set()) // first invalid frame
        set_timeout(\(T\sb\mathrm{to}\))
// upon timeout expiry: invoke recovery
\end{alltt}
    \caption{Pseudo-code for the unified receiver in TRUDI.}
    \label{fig:pseudo-code}
\end{figure}

If $V(\mathcal{A})$ is  true then the origin of the \mbox{U-frame} is authentic, and all the information it includes about keychains ($A^{g}$ tuples for which $\tau_g$ is true) are taken as valid by the receiver.
No matter the value of $\varrho_g$, the related status information $S^{g}$ is then updated as $\langle \hat{c}_g = c_g, \hat\iota_g = i_g$, $\hat\kappa_{g} = K^{g} \rangle$ and $\varrho_g$ is set to true to indicate that they are now significant.

Finally, every keychain $\mathcal{C}_{g}$ for which $\Omega_{g}$ in the U-frame is true is deleted, i.e.,  every receiver removes the mapping between $g$ and $\mathcal{C}_{g}$, making the identifier vacant. 
Contextually, it invalidates entry $S^{g}$ by setting the local variable $\varrho_g$ to false, as the keychain no longer exists.
Receiver operations are described in Fig.~\ref{fig:pseudo-code} by means of a C-like pseudo-code.

Note that, $\Tsc$ retains full control of the authentication process.
This permits to implement adaptive mechanisms where the authentication strategy changes at runtime depending on application needs and environmental conditions.
In the following, some examples are provided that show how TRUDI can deal with all the strategies proposed in the previous section.
For sake of simplicity frames will be still denoted A, J, and D, despite they actually use the common U-frame format.

\subsection{Single Keychain}
In this case $G=2$ ($\mathcal{C}$ and $\mathcal{C}'$). 
When the \gls{sc} is initialized, only $\mathcal{C}_{1}$ is set as valid in the intended recipients, by setting its state to $\langle \varrho_1 = \mathrm{true}; \hat{c}_1 = 0, \hat\iota_1 = 0, \hat\kappa_1 = K_0 \rangle$, whereas $\varrho_2= \mathrm{false}$.
The following A-frames include just one key, that initially is identified by $g=1$, whereas $\tau_2 = \mathrm{false}$.
Instead, every J-frame includes two keys.
The first \mbox{J-frame} sets  the previous keychain $\mathcal{C}_{1}$ for deletion by setting $\Omega_1=\mathrm{true}$.
Contextually, by using the values provided by $\Tsc$ in $A^{2}$, i.e., $\langle \tau_2 = \mathrm{true}; c_2 = 1, i_2 = 0, K^{2} = K_0^{'} \rangle$, the state of the new keychain $\mathcal{C}_{2}$ is initialized in receivers to $\langle \varrho_2 = \mathrm{true}; \hat{c}_2 = c_2, \hat\iota_2 = i_2, \hat\kappa_2 = K^{2} \rangle$.
Then, transmission of \mbox{A-frames} starts again, using $g=2$ as keychain identifier (i.e., $\tau_1 = \mathrm{false}$).
Finally, another \mbox{J-frame} is sent, where $A^{1} = \langle \tau_1 = \mathrm{true}; c_1 = 2, i_1 = 0, K^{1} = K_0^{''} \rangle$, which enables $\mathcal{C}_{1}$ again by setting $\langle \varrho_1 = \mathrm{true}; \hat{c}_1 = c_1, \hat\iota_1 = i_1, \hat\kappa_1 = K^{1} \rangle$ and contextually deletes $\mathcal{C}_{2}$ by setting $\Omega_2=\mathrm{true}$.
Above procedure is repeated indefinitely.
Should synchronization be lost, the receiver notifies $\Tsc$ that recovery is required.

The case when subsequent keychains overlap is similar, the only difference being that $Q$ J-frames are sent back-to-back, each one with the $\Omega$ flag set.
As soon as one of these \mbox{J-frames} is correctly received and validated, the state of the previous keychain $\mathcal{C}$ is deleted, which means that backtracking on it is no longer possible in the next J-frames.
This is not a problem, since the state of the new keychain $\mathcal{C}'$ is contextually created.

\subsection{Dual Keychain}
In this case $G=3$, as junctions for $\mathcal{C}_\mathrm{a}$ and $\mathcal{C}_\mathrm{b}$ never happen at the same time.
When the \gls{sc} is initialized, $\mathcal{C}_{1}$ and $\mathcal{C}_{2}$ are set as valid in the intended recipients by setting state variables to $\langle \varrho_1 = \mathrm{true} : \hat{c}_1 = 0, \hat\iota_1 = 0,   \hat\kappa_1 = K_0^\mathrm{a} \rangle$ and $\langle \varrho_2 = \mathrm{true}: \hat{c}_2 = 0, \hat\iota_2 = N-1, \hat\kappa_2 = K_{N-1}^\mathrm{b} \rangle$ (fully interleaved case).
Instead, $\varrho_3= \mathrm{false}$.
The following \mbox{D-frames} include authentication information for $\mathcal{C}_{1}$ and $\mathcal{C}_{2}$, while $\tau_3=\mathrm{false}$.

The first J-frame joins $\mathcal{C}_\mathrm{b}$ to $\mathcal{C}'_\mathrm{b}$.
Authentication information for them is included in $A^{2}$ and $A^{3}$, respectively.
In particular, $\langle \tau_3 = \mathrm{true}; {c}_3 = 1, i_3 = 0, K^{3}=K_0^{\mathrm{b}'} \rangle$ creates $\mathcal{C}'_\mathrm{b}$, while setting $\Omega_2$ to true deletes old $\mathcal{C}_\mathrm{b}$ and makes identifier $g=2$ vacant.
Information for $\mathcal{C}_\mathrm{a}$ is instead omitted by setting $\tau_1=\mathrm{false}$.
In this way, the related state variables on receivers are left untouched.
Subsequent \mbox{D-frames} include information for $\mathcal{C}_{1}$ ($\mathcal{C}_\mathrm{a}$) and $\mathcal{C}_{3}$ ($\mathcal{C}'_\mathrm{b}$), while $\tau_2=\mathrm{false}$.
The next J-frame joins $\mathcal{C}_\mathrm{a}$ to $\mathcal{C}'_\mathrm{a}$.
This time, authentication information for them is included in $A^{1}$ and $A^{2}$, respectively, and identifier $g=1$ is made vacant.
As can be seen, the mapping between $g$ values and keychains $\langle \mathcal{C}_\mathrm{a}, mathcal{C}_\mathrm{b} \rangle$ alternates every half sequence according to the repeating pattern $(
\langle \mathcal{C}_{1}, \mathcal{C}_{2} \rangle$, $ 
\langle \mathcal{C}_{1}, \mathcal{C}_{3} \rangle$, $ 
\langle \mathcal{C}_{2}, \mathcal{C}_{3} \rangle$, $ 
\langle \mathcal{C}_{2}, \mathcal{C}_{1} \rangle$, $ 
\langle \mathcal{C}_{3}, \mathcal{C}_{1} \rangle$, $ 
\langle \mathcal{C}_{3}, \mathcal{C}_{2} \rangle$, $ 
\langle \mathcal{C}_{1}, \mathcal{C}_{2} \rangle$, $ 
\ldots )$.

The sparsely-interleaved dual-keychain case works the same way for what concerns J-frames, but some differences exist because a mixture of A-frames and D-frames is used in between.

\section{Conclusions}
\label{sec:Conc}
Guaranteeing networked system security is becoming crucial also in the automotive and embedded scenarios.
To this aim both Ethernet and CAN XL define standard security countermeasures based on symmetric cryptography, which ensure message integrity and confidentiality.
Nevertheless, neither of these solutions is able to guarantee receivers that the origin of a multicast frame is authentic, which makes them weak against masquerade attacks.
By leveraging \gls{ecu} software vulnerabilities in connected cars, large-scale coordinated attacks can be delivered, which may involve many thousands of vehicles at the same time, with catastrophic consequences.

In this paper some robust multicast origin authentication strategies are described, which stem from TESLA with immediate delivery and exploit single-overlapped or dual-interleaved keychains.
These strategies effectively meet the requirements of small real-time networked embedded systems that rely on multidrop busses and unmanaged switches, and are connected to the Internet as well, as modern vehicles.
Unless they are physically tampered (in which case the attacker can do literally everything), such systems do not suffer from MITM attacks.

As expected, a trade-off exists among the achievable security and reliability levels, as well as resource consumption (processing and communication).
As a realistic example, a sparse dual-interleaved keychain with $n=1023$ and $m=31$ requires two keychains with $1024$ keys each.
Key transmission efficiency is $\eta_\mathrm{KT}=0.96875$ and, if MD5 is adopted, memory requirements on the transmitter amount to (at least) $\SI{32}{kB}$.
If used to convey a cyclic stream with period $\SI{10}{ms}$, key disclosure never occurs more than $\SI{0.31}{s}$ after commitment.
Any error bursts affecting up to $511$ adjacent frames is tolerated (i.e., a communication outage lasting up to $\SI{5}{s}$ does not enforce recovery), and $16$ evenly-spaced synchronization opportunities are provided in the start-up phase.

In addition, a unified receiver architecture has been proposed, named TRUDI, which moves all details about the way origin-authentication is carried out (that impact on performance and requirements) on the transmitter side, enabling authentication to adapt at runtime to dynamically changing environmental conditions (e.g., failure rate on the bus).

Future work will include a  stable definition for TRUDI, including its frame format, and a thorough analysis of its performance and effectiveness against different types of attacks.
Moreover, we will investigate how to include its basic concepts directly in MACsec and CANsec, even though doing so breaks backward compatibility.

\bibliographystyle{ieeetr}
\bibliography{biblio}

\begin{IEEEbiography}[{\includegraphics[width=1in,height=1.25in,clip,keepaspectratio]{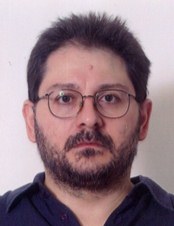}}]{Gianluca Cena}(SM’09) received the M.S. degree in electronic engineering and the Ph.D. degree in information and system engineering from the Politecnico di Torino, Italy, in 1991 and 1996, respectively. Since 2005 he has been a Director of Research with the National Research Council of Italy (CNR-IEIIT). His research interests include wired and wireless industrial communication systems, real-time protocols, and automotive networks. In these areas he has co-authored about 170 technical papers and one international patent. He received the Best Paper Award of the IEEE TRANSACTIONS ON INDUSTRIAL INFORMATICS in 2017 and of the IEEE Workshop on Factory Communication Systems in 2004, 2010, 2017, 2019, and 2020. Dr. Cena served as a Program Co-Chairman of the IEEE Workshop on Factory Communication Systems in 2006 and 2008. Since 2009 he has been an Associate Editor of the IEEE TRANSACTIONS ON INDUSTRIAL INFORMATICS.
\end{IEEEbiography}

\begin{IEEEbiography}[{\includegraphics[width=1in,height=1.25in,clip,keepaspectratio]{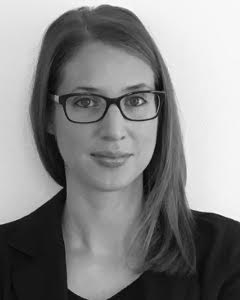}}]{Lucia Seno} received the B.S. and M.S. degrees  in Automation Engineering and the Ph.D. degree in Information Engineering from the University of Padova, Padova, Italy, in 2004, 2007, and 2011, respectively.
She has been a Researcher with the Institute of Electronics, Computer, and Telecommunication Engineering of the National Research Council of Italy in Padova, and, subsequently, in Torino, Italy, from 2011 to 2023. Currently she is with the Top Enterprise Market Pre Sales at TIM SpA. Her interests include industrial communication systems and technologies, wireless networks and wireless sensor networks for real-time and safety-critical control applications, formal verification of system security and communication protocols, and model-checking. 
\end{IEEEbiography}

\begin{IEEEbiography}[{\includegraphics[width=1in,height=1.25in,clip,keepaspectratio]{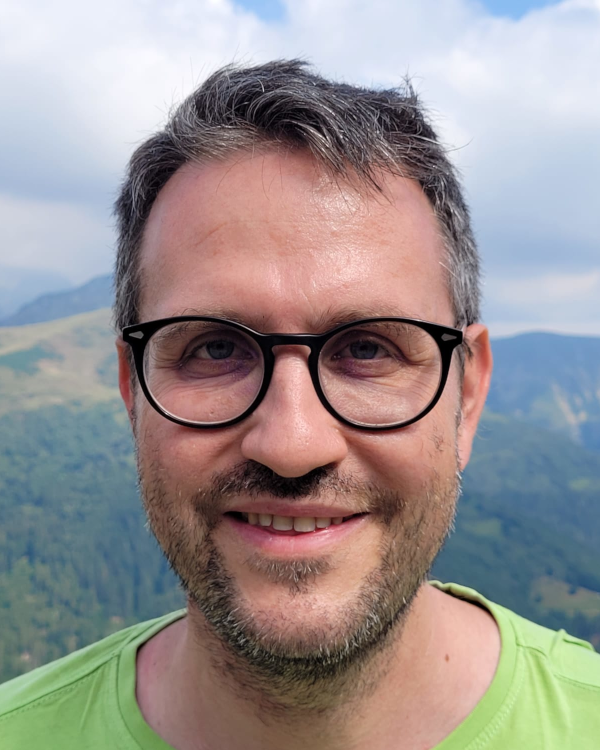}}]{Stefano Scanzio} (S’06-M’12-SM’22) received the Laurea and Ph.D. degrees in computer science from Politecnico di Torino, Turin, Italy, in 2004 and 2008, respectively.
From 2004 to 2009, he was with the Department of Computer Engineering, Politecnico di Torino, where he was involved in research on speech recognition and classification methods and algorithms. Since 2009, he has been with the National Research Council of Italy, where he is currently a Senior Researcher with the institute CNR-IEIIT. He teaches several courses on computer science with Politecnico di Torino and Università degli Studi di Pavia. He has authored or coauthored more than 100 papers in international journals and conferences, in the areas of industrial communication systems, real-time networks, wireless networks, and artificial intelligence. He took part in the program and organizing committees of many international conferences of primary importance in his research areas. He received the 2017 Best Paper
Award from IEEE TRANSACTIONS ON INDUSTRIAL INFORMATICS, and the Best Paper Awards for three papers presented at the IEEE Workshops on Factory Communication Systems, in 2010, 2017, and 2019, and for a paper presented at the IEEE International Conference on Factory Communication Systems, in 2020. He is an Associate Editor of IEEE ACCESS, Ad Hoc Networks (Elsevier), and Electronics (MDPI).
\end{IEEEbiography}

\vfill

\end{document}